\documentclass[11pt,twoside]{article}
\usepackage{./asp2014}

\aspSuppressVolSlug
\resetcounters

\begin{document}

\title{NuSTAR observations of the Dwarf Nova GK Persei in 2015: comparison between outburst and quiescent phases}
\author{Yuuki Wada,$^1$ Takayuki Yuasa, Kazuhiro Nakazawa,$^1$ Kazuo Makishima,$^2$ Takayuki Hayashi,$^{3,4}$ and Manabu Ishida$^5$
\\
\affil{$^1$ Graduate school of Science, The University of Tokyo, Bunkyo, Tokyo, Japan; \email{wada@juno.phys.s.u-tokyo.ac.jp}}
\affil{$^2$ RIKEN, Wako, Saitama, Japan;}
\affil{$^3$ Nagoya University, Nagoya, Aichi, Japan; \email{}}}
\affil{$^4$ GSFC/NASA, Greenbelt, MD, USA; \email{}}
\affil{$^5$ ISAS/JAXA, Sagamihara, Kanagawa, Japan; \email{}}

\paperauthor{Yuuki Wada}{wada@juno.phys.s.u-tokyo.ac.jp}{}{The University of Tokyo}{Graduate School of Science}{Bunkyo}{Tokyo}{113-0033}{Japan}
\paperauthor{Takayuki Yuasa}{}{}{}{}{}{}{}{}
\paperauthor{Kazuhiro Nakazawa}{}{}{The University of Tokyo}{Graduate School of Sciemce}{Bunkyo}{Tokyo}{113-0033}{Japan}
\paperauthor{Kazuo Makishima}{}{}{Riken}{MAXI team}{Wako}{Saitama}{Postal Code}{Japan}
\paperauthor{Takayuki Hayashi}{}{}{Nagoya University}{}{Nagoya}{Aichi}{Postal Code}{Japan}
\paperauthor{Manabu Ishida}{}{}{JAXA}{}{Sagamihara}{Kanagawa}{Postal Code}{Japan}

\begin{abstract}
We report on {\it NuSTAR} observations of the Intermediate Polar GK Persei which also behaves as a Dwarf Nova. It exhibited a Dwarf Nova outburst in 2015 March-April. The object was observed in 3--79 keV X-rays with {\sl NuSTAR}, once at the outburst peak, and again in 2015 September during quiescence. The 5-50 keV flux during the outburst was 26 times higher than that during the quiescence. With a multi-temperature emission model and a reflection model, we derived the post-shock temperature as 
$19.2 \pm 0.7$ keV in the outburst, and $38.5 ^{+4.1}_{-3.6}$ keV in the quiescence. This temperature difference is considered to reflect changes in the radius at which the accreting matter, forming an accretion disk, is captured by the magnetosphere of the white dwarf (WD). Assuming that this radius scales as the power of -2/7 of the mass accretion rate, and utilizing the two temperature measurements, as well as the standard mass-radius relation of WDs, we determined the WD mass in GK Persei as $0.90 \pm 0.06$ solar masses. The magnetic field is estimated as $4 \times 10^{5}$ G.
\end{abstract}

\section{Introduction}
 Cataclysmic Variables (CVs) are close binary systems consisting of a mass-accreting white dwarf (WD) and a mass-donating late-type main-sequence companion star. Intermediate Polars (IPs), an important subclass of CVs, are considered to involve WDs with magnetic field of $B = 10^{5-6}$ G. In an IP, gas from the companion forms an accretion disk down to a radius $R_{\rm in}$ where gravity is counter-balanced by the magnetic pressure. Then the gas is captured by the WD's magnetosphere, and accretes onto the WD surface to form accretion columns due to strong magnetic field. In the accretion column, the gas is heated up to $10^{7-8}$ K by a standing shock, and lands onto the WD surface emitting thermal X-rays. If $R_{\rm in}$ is far enough from the WD surface, the temperature $T_{\rm s}$ just below the shock is proportional to gravitational potential of the WD (Aizu 1973):
\begin{equation}
T_{\rm s} = \frac{3}{8} \mu m_{\rm H} \frac{GM_{\rm WD}}{R_{\rm WD}},
\end{equation}
where $\mu$ is mean molecular weight and $m_{\rm H}$ is the proton mass. Therefore the WD can be estimated by combining the measured $T_{\rm s}$ with the standard mass v.s. radius ($M_{\rm WD}$-$R_{\rm WD}$) relation (Nauenberg 1972).

An X-ray spectrum from IPs is a superposition of optically-thin thermal emissions of various temperatures, from $T_{\rm s}$ downwards. To determine $T_{\rm s}$, it is hence important to accurately measure both the hard X-ray continuum, and the ratio of Fe XXV and XXVI lines at $\sim 7$ keV. This is because the hard X-ray continuum is sensitive to the hottest components (with temperature $\sim T_{\rm s}$), whereas the latter tells us contributions from cooler components arising closer to the WD surface. This method has been established with X-ray satellites such as {\sl Ginga} (e.g. Ishida 1991), {\sl ASCA} (e.g. Fujimoto \& Ishida 1997, Ezuka \& Ishida 1997), {\sl RXTE} (e.g. Suleimanov et al. 2005), {\sl INTEGRAL} (e.g. Falanga et al. 2005), {\sl Swift} (e.g. Brunschweiger et al. 2009), and {\sl Suzaku} (e.g. Yuasa et al. 2010, Hayashi et al. 2011). 

GK Persei, a Cataclysmic Variable system, is interestingly categorized as both a Dwarf Nova and an IP. After a classical Nova explosion in 1901 (Williams 1901; Hale 1901), it repeats, every 2--3 years, Dwarf Nova outbursts each lasting for 2 months (e.g. \v{S}imon 2001). During outbursts, the optical and X-ray luminosities increase by a factor of 10--20. The distance was measured to be $477 ^{+28}_{-25}$ pc (Harrison \& Bornak 2013). Reinsch (1994) and Morales-Rueda et al. (2002) optically determined the WD mass as $M_{\rm WD} \geq 0.78 M_{\rm \odot}$ and $M_{\rm WD} \geq 0.55 M_{\rm \odot}$, respectively. 

An outburst from GK Persei started in March 2015, and continued for 2 months (Wilber et al. 2015). During this outburst, a ToO (Target of Opportunity) observation of {\sl NuSTAR} was triggered (Zemko et al. 2016). Suleimanov et al. (2016) analyzed the ToO data and constrained the WD mass as $M_{\rm WD} = 0.86 \pm 0.02 M_{\rm \odot}$. {\sl Suzaku} was serendipitously pointed at the onset of this outburst (Yuasa et al. 2016). With {\sl NuSTAR}, we observed GK Persei again, after the object returned to its quiescence. The present paper describes a combined analysis of the outburst and quiescence data from {\sl NuSTAR}.

\section{Observation and Data Reduction}
The X-ray satellite {\sl NuSTAR} (Harrison et al. 2013), launched on 2012 June 13, has a 3--79 keV sensibility. It is thus suited for our purpose, because we can study both the hard X-ray continuum and the Fe-K lines.

As described in \S1, GK Persei was observed twice with {\sl NuSTAR}, once at the outburst peak and the other after the outburst terminated. Details of the observations are shown in table 1. We extracted the data from the two focal plane modules (FPMA and FPMB) with {\tt nupipeline} and {\tt nuproducts} included in NUSTARDAS version 1.5.1. The source region was chosen to be a circle of $150''$ radius for the outburst data, and $80''$ for the quiescence data. The remaining region was used for background. We used NuSTAR CALDB version 20150904. The X-ray spectra were analyzed with XSPEC version 12.9.0. (Arnaud 1996).
\begin{table}[t!]
\begin{center}
\caption{Observation log of GK Persei by {\it NuSTAR}}
\begin{tabular}{cccccc}
\hline
			& observation ID	& start		& stop		& exposure	& PI			\\ \hline
outburst 		& 90001008002	& 2015-04-04	& 2015-04-06	& 42 ks 		& ToO		\\
			&				& 02:46:07	& 15:10:35	& 			& 			\\ \hline
quiescence 	& 30101021002	& 2015-09-08	& 2015-09-11	& 72 ks 		& T. Yuasa	\\
			&				& 15:46:08 	& 02:04:09 	&			&			\\ \hline
\end{tabular}
\end{center}  
\end{table}

\section{Results}
\subsection{Outburst and quiescent spectra}
Figure 1 shows 3--50 keV spectra of the outburst and quiescence data sets. The background has been subtracted, but the instrumental response has not been removed. Data from FPMA and FPMB are separately plotted. As reported by Zemko et al. (2016) and Suleimanov et al. (2016), the hard X-ray continuum was detected up to 50 keV during the outburst. In addition, we obtained, for the first time, high-quality hard X-ray data of this object in quiescence. The 3-50 keV count rate of FPFA plus FPMB was 18.09 $\pm$ 0.02 count s$^{-1}$ in outburst, and 1.080 $\pm$ 0.006 count s$^{-1}$ in quiescence. Thus, the outburst data have 17.5 times higher count rate than quiescence. The spectra, particularly the outburst data, exhibit Fe-K line complex at $\sim$ 6.4 keV. From this energy, we regard the lines as mainly of fluorescence origin (from the WD surface and the accreting matter), rather than ionized lines from the accretion columns.

In the bottom panel, we show the ratio between the two spectra. It reveals three features of the outburst spectrum, in comparison with that in quiescence. Namely, a stronger low-energy absorption, the stronger Fe-K line, and most importantly, a lower value of $T_{\rm s}$ as evidenced by a concave shape peaking at $\sim$ 20 keV.
\articlefigure[width=0.5\textwidth]{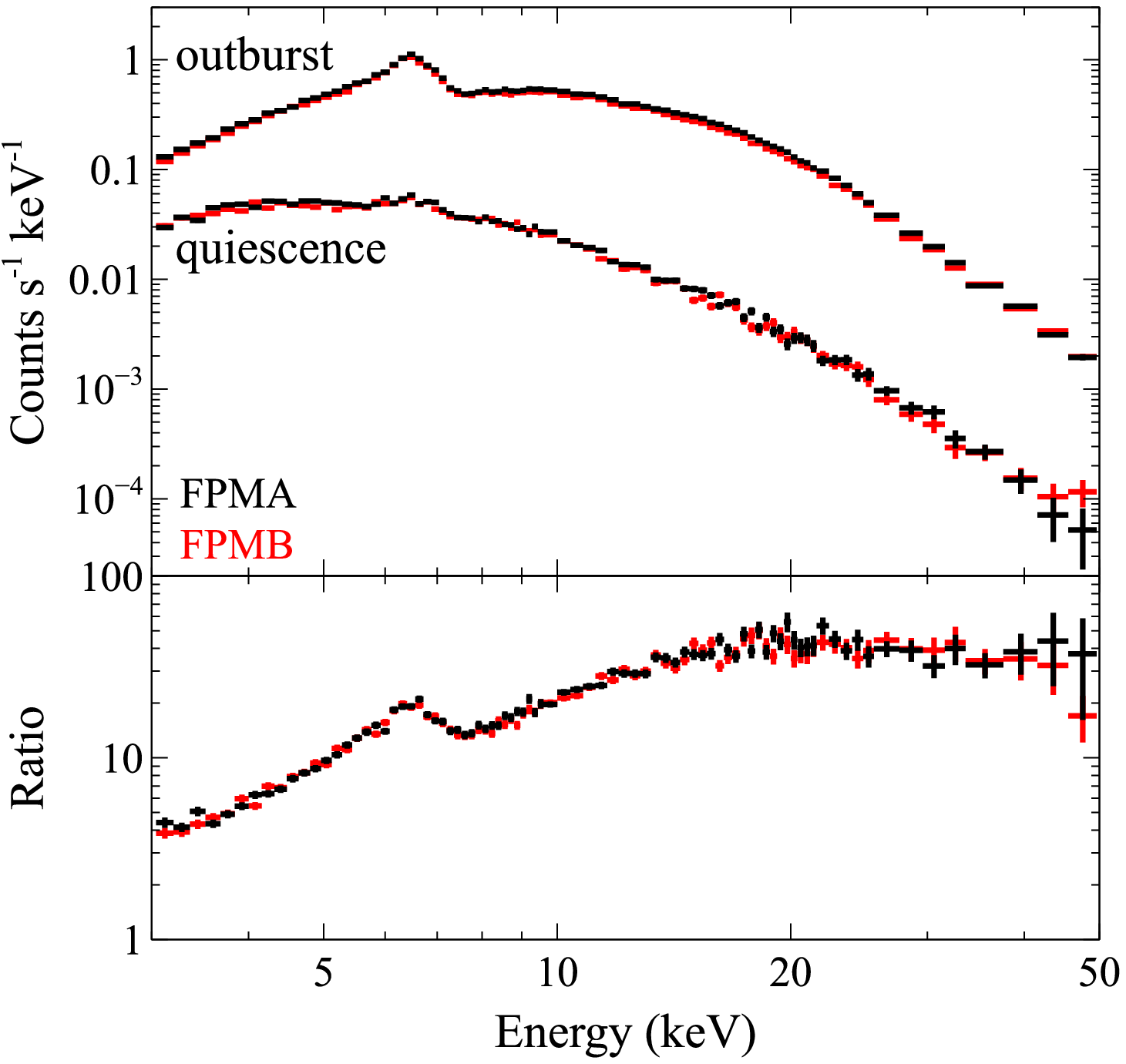}{spectral ratio}{The {\it NuSTAR} spectra of GK Persei, obtained in an outburst and quiescence. The black and red indicate data from FPMA and FPMB, respectively. The outburst-to-quiescence spectral ratio is shown in the bottom panel.} 

\subsection{Spectral fitting}
In order to quantify the spectral properties (including the difference between the two), we proceed to fit the two spectra with a common physical model over the 5--50 keV range. To reproduce the expected multi-temperature optically-thin plasma emission from the accretion column, we utilized {\tt cemekl} model (Done \& Osborne 1997), which superposes a plasma emission code called {\tt mekal} model of multiple temperatures. The differential emission measure is defined in the {\tt cemekl} model as $d{\rm (EM)} \propto (T / T_{\rm s})^{\alpha -1} dT$,
where $\alpha$ is a positive parameter. Suleimanov et al. (2005) \& Falanga et al. (2005) predicted $\alpha = 0.43$ by computating pressure gradient and gravity effects in the accretion column. We hence employed this value. In addition, photo-absorption model called {\tt phabs}, {\tt reflect} model (Magdziarz \& Zdziarski 1995) to calcurate reflection component from the WD surface, and a gaussian for the Fe-K$\alpha$ line were added. In fitting the outburst spectrum, a partial covering model {\tt pcfabs} was also added to represent a condition wherein a fraction of the continuum reaches us through a thick absorbing matter. Assuming that the standing shock is located almost on the WD surface, we fix the solid angle of the reflection matter to $2\pi$. For the outburst spectrum, 1\% systematic error was included.

The spectra and best-fit models are shown in Figure 2, and their parameters are summarized in table 2. Errors are at 90\% confidence level. The reduced $\chi^{2}$ was 1.11 with 410 degrees of freedom in outburst, and 1.14 with 135 degrees of freedom in quiescence. Both fittings were acceptable. The shock temperature was obtained as $T_{\rm s} = 19.2 \pm$ 0.07 keV in outburst, and 38.5 $^{+4.1}_{-3.5}$ keV in quiescence. Thus, $T_{\rm s}$ almost halved as the accretion rate increased. The 5-50 keV absorbed flux in outburst was 26 times higher than that in quiescence.

For our purpose, we need to calculate the bolometric flux $F$, which is thought to be directly proportional to the mass accretion rate $\dot{M}$. For both spectra, $F$ was derived in the folllowing way, starting from the absorbed 5--50 keV flux. First, we removed the overall and partial absorption, and the reflection. Second, the flux above 50 keV was included by extrapolating the best-fit model up to 100 keV. Finally, the contribution below 5 keV was incorporated by integrating the best-fit model down to 0.01 keV. The flux above 100 keV and below 0.01 keV are both estimated to be $\ll 0.01 F$. As show in Table 2, the difference in $F$ between the two spectra amounts to a factor of 74.

\begin{table}[t!]
\begin{center}
\caption{The obtained parameters from the best-fit model.}
\begin{tabular}{cccc}
\hline
			& $T_{\rm s}$			& flux (5--50 keV)					& $F$					\\
			& keV				& erg cm$^{-2}$ s$^{-1}$				& erg cm$^{-2}$ s$^{-1}$				\\ \hline
outburst 		& 19.2 $\pm$ 0.07		& $(6.6^{+0.8}_{-0.7}) \times 10^{-10} $	& $(3.5 \pm 0.4) \times 10^{-9}$		\\
quiescence 	& 38.5 $^{+4.1}_{-3.5}$	& $(2.6 \pm 0.6) \times 10^{-11} $		& $(4.7 \pm 1.1) \times 10^{-11}$		\\ \hline
\end{tabular}
\end{center}  
\end{table}

\articlefiguretwo{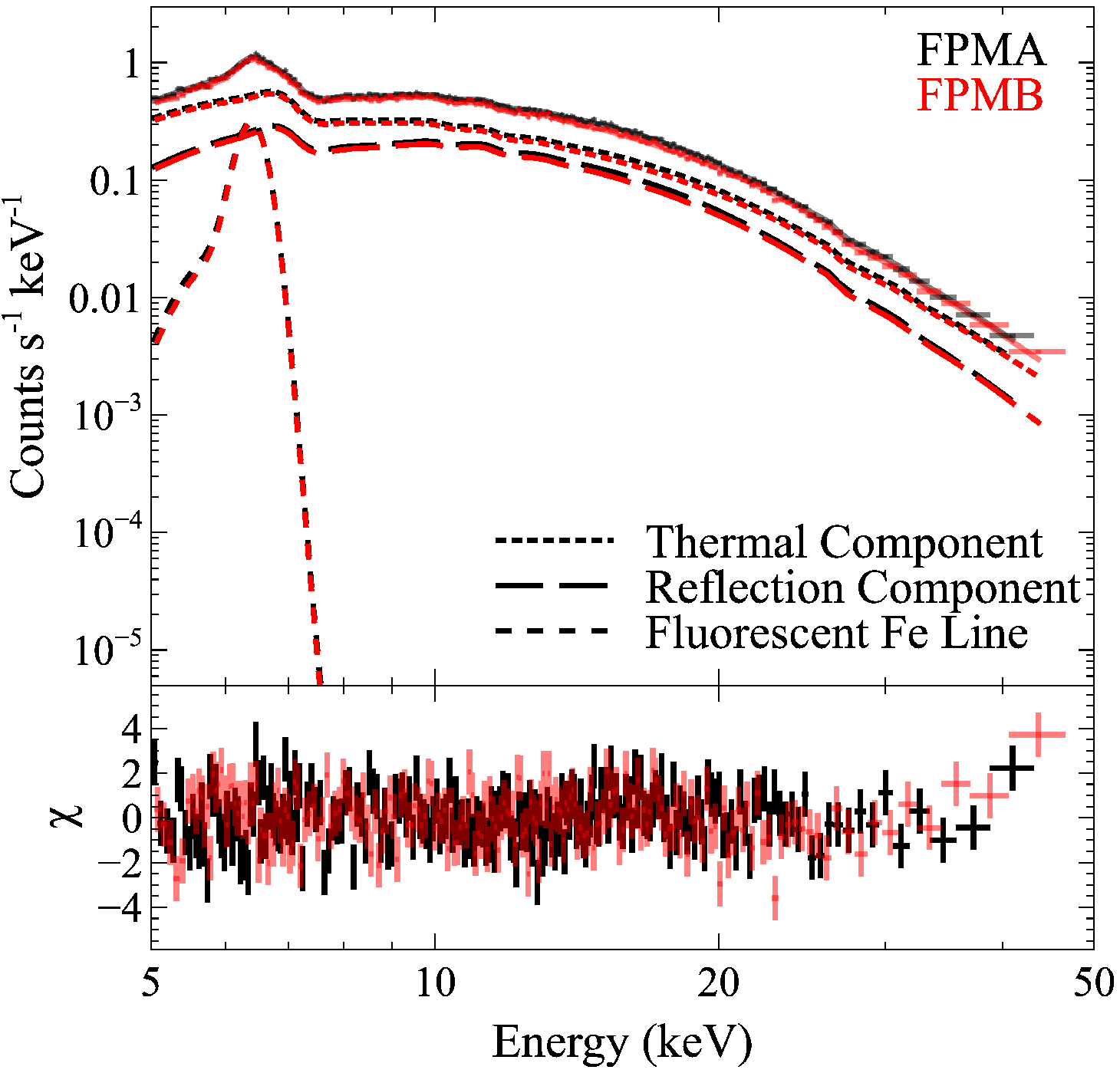}{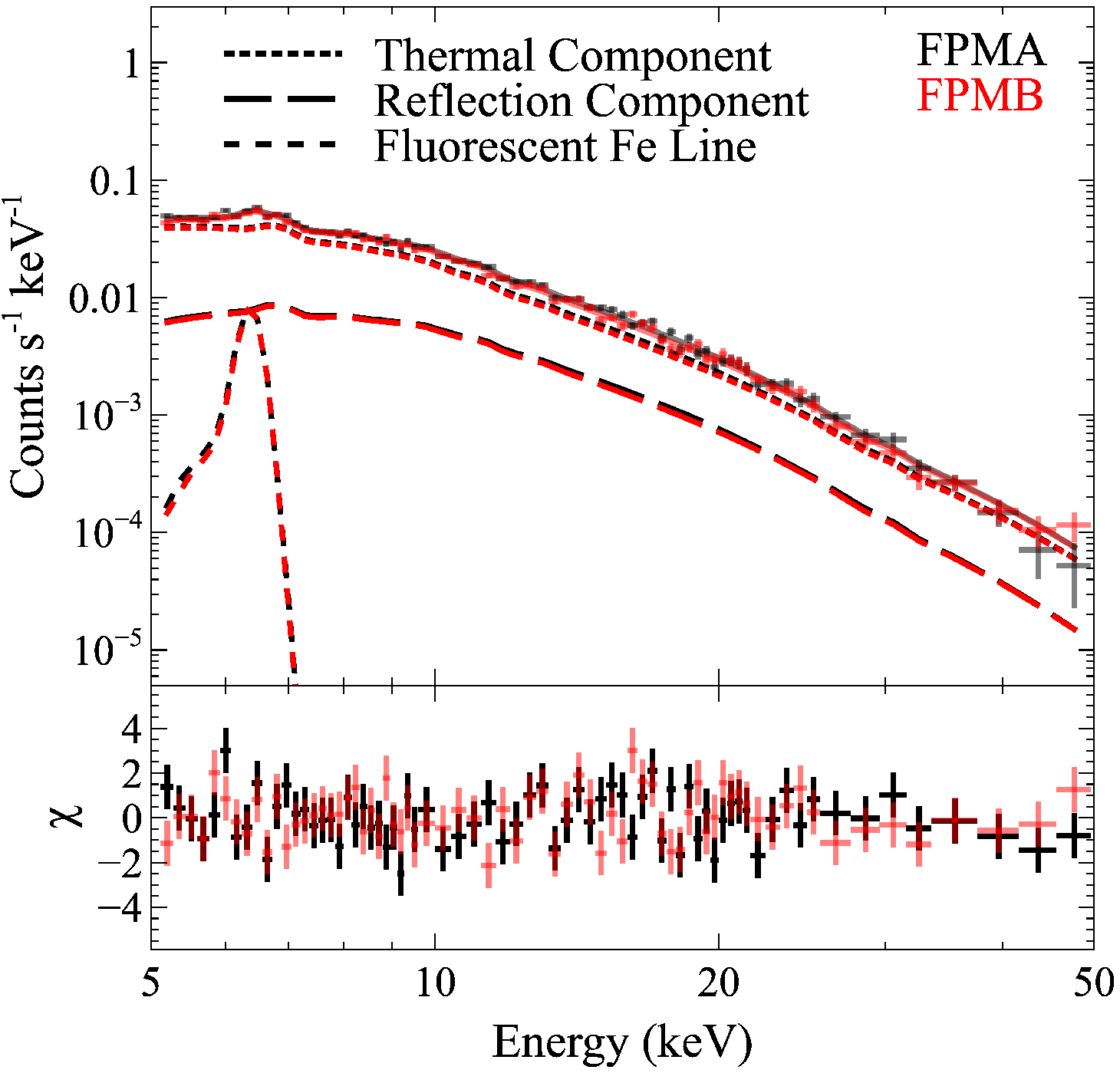}{figure2}{Results of the model fitting to the outburst (left panel) and the quiescent (right panel) spectrum.
The best-fit models and their components are overlaid on the spectra. The plot color is the same as in figure 2.}

\section{Discussion}
\subsection{The origin of $T_{\rm s}$ difference}
As described in \S 1, the accreting matter in an IP source is considered to be captured by the magnetic field at $R_{\rm in}$, where it starts the free-fall motion. When $R_{\rm in}$ is close to the WD surface, eq. (1) is modified as
\begin{equation}
T_{\rm s} = \frac{3}{8} \mu m_{H} \frac{GM_{\rm WD}}{R_{\rm WD}} \left( 1-\frac{R_{\rm WD}}{R_{\rm in}} \right).
\end{equation}
According to Ghosh \& Lamb (1979), $R_{\rm in}$ is determined by an equilibrium between the gravitational pull working on the gas and the outward magnetic pressure, and is described as
\begin{equation}
\frac{R_{\rm in}}{R_{\rm WD}} \approx 2.3 \: M_{\rm WD}^{-1/7} \: R_{\rm WD}^{5/7} \left(\frac{\dot{M}}{10^{20} \: {\rm g \: s^{-1}}}\right)^{-2/7} \left( \frac{B}{10^{6} \: {\rm G}} \right)^{4/7},
\end{equation}
where $B$ means the magnetic strength on the WD surface. This equation indicates that $R_{\rm in}$ shrinks when $\dot{M}$ increases. Therefore $T_{\rm s}$ must be lower in outburst than in quiescence, in agreement with our observation.

\subsection{The WD mass estimation}
To estimate the WD mass accurately, we need to estimate $R_{\rm in}$. From eq. (3), the ratio of $R_{\rm in}$ in outburst and quiescence, denoted $\gamma$, is derived as
\begin{equation}
\gamma = \frac{R_{\rm in}^{\rm qui}}{R_{\rm in}^{\rm out}} = \left( \frac{\dot{M}_{\rm qui}}{\dot{M}_{\rm out}} \right)^{-2/7}.
\end{equation}
Considering the energy release from $R_{\rm in}$ to WD surface, $\dot{M}$ must be related to the bolometric luminosity $L$ as
\begin{equation}
\dot{M}=L \times \left( \frac{GM_{\rm WD}}{R_{\rm WD}} \right)^{-1} \left( 1-\frac{R_{\rm WD}}{R_{\rm in}} \right)^{-1}.
\end{equation}
From eq. (2), (4), and (5), $\gamma$ can be derived with two sets of $T_{\rm s}$ and $F$ as
\begin{equation}
\gamma=\left\{ \frac{L_{\rm q} \times T_{\rm s}^{\rm out}}{L_{\rm o} \times T_{\rm s}^{\rm qui}} \right\}^{-2/7}=\left\{ \frac{F_{\rm q} \times T_{\rm s}^{\rm out}}{F_{\rm o} \times T_{\rm s}^{\rm qui}} \right\}^{-2/7}.
\end{equation}
Thus, $\gamma$ was calculated as $4.2 \pm 0.7$ with the obtained $T_{\rm s}$ and $F$.

Given eq.(6), let us proceed to the $M_{\rm WD}$ determination. When we specify a value of $M_{\rm WD}$, we can derive the associated $R_{\rm WD}$ from the $M_{\rm WD}$-$R_{\rm WD}$ relation, and utilize eq. (1) to calculate $T_{\rm s}$ for $R_{\rm in} \rightarrow \infty$, to be denoted as $T_{\rm s}^{\infty}$. These values are shown in Table 3. Then $R_{\rm in}^{\rm out}$ and $R_{\rm in}^{\rm qui}$ can be derived by comparing this $T_{\rm s}^{\infty}$ with two measurements of $T_{\rm s}$ using eq. (2). These values, together with the ratio $R_{\rm in}^{\rm qui}/R_{\rm in}^{\rm out}$, are also given in Table 3. Finally, by requiring this ratio to coincide with $\gamma$ of eq.(6), the WD mass can be determined as $M_{\rm WD} = 0.90 \pm 0.06 \: M_{\odot}$. These results can be substituted to obtain $R_{\rm in}^{\rm qui} \approx 7.3 \: R_{\rm WD}$, $R_{\rm in}^{\rm out} \approx 1.8 \: R_{\rm WD}$, $B \sim 4 \times 10^{5}$ G. 

In actual calculation, we have to take the shock height in account. The parameters shown in Table 3 were derived by using eq. (6) of Suleimanov et al. (2016), which includes the effect of the shock height changes. The calculated $T_{\rm s}^{\infty}$ with the Suleimanov's equation were slightly lower than that with eq. (1). 

\begin{table}[t!]
\begin{center}
\caption{Parameters calculated with the observed $T_{\rm s}^{\rm out}$ and $T_{\rm s}^{\rm qui}$ as a function of $M_{\rm WD}$.}
\begin{tabular}{cccccc}
\hline
$M_{\rm WD}/M_{\odot}$	& $R_{\rm WD}/10^{-2} \: R_{\odot}$	& $T_{\rm s}^{\infty}$ (keV)	& $R_{\rm in}^{\rm out}$/$R_{\rm WD}$		& $R_{\rm in}^{\rm qui}$/$R_{\rm WD}$		& $R_{\rm in}^{\rm qui}/R_{\rm in}^{\rm out}$	\\ \hline
0.7					& 1.1							& 27.9					& 3.2						& ($T_{\rm s}^{\rm qui} > T_{\rm s}^{\infty}$)		& --		\\
0.8					& 1.0							& 35.5					& 2.2						& ($T_{\rm s}^{\rm qui} > T_{\rm s}^{\infty}$)		& --		\\
0.9					& 0.90						& 44.9					& 1.7						& 7.1										& 4.1		\\
1.0					& 0.80						& 57.1					& 1.5						& 3.1										& 2.1		\\
1.1					& 0.67						& 73.3					& 1.4						& 2.1										& 1.6		\\ \hline
\end{tabular}
\end{center}
\end{table}



\end{document}